\documentclass[fleqn,usenatbib]{mnras}

\usepackage{newtxtext,newtxmath}

\usepackage[T1]{fontenc}
\usepackage{ae,aecompl}

\usepackage{graphicx}	
\usepackage{amsmath}	
\usepackage{amssymb}	

\usepackage{hyperref}

\newcommand{\revision}[1]{#1}
\newcommand{\kms}{{~\rm km\; s^{-1}}}

\newcommand{\cm}{{~\rm cm}}
\newcommand{\s}{{~\rm s}}

\newcommand{\K}{{~\rm K}}
\newcommand{\erg}{{~\rm erg}}

\newcommand{\days}{{~\rm days}}
\newcommand{\Rsun}{{~\rm R_{\sun}}}
\newcommand{\Msun}{{~\rm M_{\sun}}}

\newcommand{\erggs}{{~\rm erg~g^{-1}~s^{-1}}}

\newcommand{\cmg}{{~\rm cm^2~g^{-1}}}
\newcommand{\As}{\ensuremath{A_{\rm s}}}
\newcommand{\DMdiff}{\ensuremath{\Delta M_{\rm diff}}}
\newcommand{\Dtexp}{\ensuremath{\Delta t_{\rm exp}}}
\newcommand{\erfc}{\ensuremath{\rm erfc}}
\newcommand{\fcomp}{\ensuremath{f_{\rm comp}}}
\newcommand{\fdom}{\ensuremath{f_{\rm DOM}}}
\newcommand{\Ldirect}{\ensuremath{L_{\rm direct}}}
\newcommand{\Ltail}{\ensuremath{L_{\rm tail}}}
\newcommand{\Mdom}{\ensuremath{M_{\rm DOM}}}
\newcommand{\Mej}{\ensuremath{M_{\rm ej}}}
\newcommand{\MNi}{\ensuremath{M_{\rm Ni}}}
\newcommand{\Ms}{\ensuremath{M_{\rm s}}}
\newcommand{\nickel}{{\ensuremath{^{56}\mathrm{Ni}}}}
\newcommand{\pf}{\ensuremath{p_{\rm f}}}
\newcommand{\ps}{\ensuremath{p_{\rm s}}}
\newcommand{\rdiff}{\ensuremath{r_{\rm diff}}}
\newcommand{\rhof}{\ensuremath{\rho_{\rm f}}}
\newcommand{\rhos}{\ensuremath{\rho_{\rm s}}}
\newcommand{\rph}{\ensuremath{r_{\rm ph}}}
\newcommand{\rs}{\ensuremath{r_{\rm s}}}
\newcommand{\Teff}{\ensuremath{T_{\rm eff}}}
\newcommand{\tdiff}{\ensuremath{t_{\rm diff}}}
\newcommand{\ts}{\ensuremath{t_{\rm s}}}
\newcommand{\vdiff}{\ensuremath{v_{\rm diff}}}
\newcommand{\vdom}{\ensuremath{v_{\rm DOM}}}
\newcommand{\ve}{\ensuremath{v_{\rm e}}}
\newcommand{\vs}{\ensuremath{v_{\rm s}}}
\newcommand{\vsnorm}{\ensuremath{\frac{\vs}{\ve\fcomp}}}

\title[Early UV emission from DOM in SNe~Ia]{Early UV emission from disk-originated matter (DOM) in type Ia supernovae in the double degenerate scenario}

\author[Levanon \& Soker]{
Naveh Levanon\thanks{E-mails: \href{mailto:nlevanon@campus.technion.ac.il}{nlevanon@campus.technion.ac.il},	\href{mailto:soker@physics.technion.ac.il}{soker@physics.technion.ac.il}}
and Noam Soker\footnotemark[1]
\\
Department of Physics, Technion -- Israel Institute of Technology, Haifa 32000 Israel
}

\date{Accepted XXX. Received YYY; in original form ZZZ}

\pubyear{2017}

\begin{document}
\label{firstpage}
\pagerange{\pageref{firstpage}--\pageref{lastpage}}
\maketitle

\begin{abstract}
We show that the blue and UV excess emission at the first few days of some type Ia supernovae (SNe~Ia) can be accounted for in the double degenerate (DD) scenario by the collision of the SN ejecta with circumstellar matter that was blown by the accretion disk formed during the merger process of the two white dwarfs (WDs).
We assume that in cases of excess early light the disk blows the circumstellar matter, that we term disk-originated matter (DOM), hours to days before explosion.
To perform our analysis we first provide a model-based definition for early excess light, replacing the definition of excess light relative to a power-law fit to the rising luminosity. 
We then examine the light curves of the SNe~Ia iPTF14atg and SN~2012cg, and find that the collision of the ejecta with a DOM in the frame of the DD scenario can account for their early excess emission.
Thus, early excess light does not necessarily imply the presence of a stellar companion in the frame of the single-degenerate scenario.  
Our findings further increase the variety of phenomena that the DD scenario can account for, and emphasize the need to consider all different SN~Ia scenarios when interpreting observations.
\end{abstract}




\section{Introduction}
\label{sec:introduction}

There is no consensus on the leading progenitor scenario for Type Ia
supernova (SN~Ia) explosions.
There are six scenarios that try to explain how carbon-oxygen white
dwarfs (WDs) evolve to thermonuclear detonation.
We list them here by their alphabetical order.
A comparison between the first five various scenarios is given by
\citet{Tsebrenko2015} and \citet{Soker2015}.
\revision{They also discuss the strong points and difficulties of the scenarios, and the properties that distinguish them from each other.}
(a) \textit{The core-degenerate (CD) scenario}
\citep[e.g.][]{Livio2003, Kashi2011, Soker2011, Ilkov2012,
Ilkov2013, Soker2013, Aznar-Siguan2015},
where during common envelope evolution a WD merges with the hot core of a
massive asymptotic giant branch (AGB) star.
Explosion might occur shortly, even before expelling the entire common envelope, or a long time after merger, up to billions of years.
(b) \textit{The double degenerate (DD) scenario},
where explosion occurs after two WDs merge \citep[e.g.][]{Webbink1984, Iben1984}.
It is not clear in this scenario how long after merger explosion occurs \citep[e.g.][]{Loren-Aguilar2009, vanKerkwijk2010, Pakmor2013, Levanon2015}.
(c) \textit{The 'double-detonation' (DDet) mechanism}
\citep[e.g.][]{Woosley1994, Livne1995, Shen2013},
where a sub-Chandrasekhar mass WD accumulates a layer of helium-rich material on its surface.
The CO WD is ignited to explosion by the detonation of the helium layer. 
(d) \textit{The single degenerate (SD) scenario}
\citep[e.g.][]{Whelan1973, Nomoto1982, Han2004},
where a non-degenerate stellar companion transfers mass to the WD.
The WD explodes when it reaches a mass about equal to the Chandrasekhar mass limit.
(e) \textit{The WD-WD collision (WWC) scenario}
\citep[e.g.][]{Raskin2009, Rosswog2009, Loren-Aguilar2010, Thompson2011, Kushnir2013, Aznar-Siguan2013, Aznar-Siguan2014},
where two WDs collide and immediately ignite the nuclear reaction that powers the explosion.
(f) \textit{The singly-evolved star (SES) scenario},
where pycnonuclear reactions drive powerful detonations in single CO white dwarfs \citep{Chiosi2015}.
The SES scenario is a relatively undeveloped scenario.
\citet{Papish2015} list two problems with the SES scenario.

Several papers point to excess luminosity in early ($\la 5 \days$) SN~Ia light curves as evidence for a collision of the SN ejecta with a companion in the frame of the SD scenario.
This is claimed for both the normal SN~Ia event SN~2012cg \citep{Marion2016} and a peculiar SN~2002es-like event, iPTF14atg \citep{Cao2015}.
In this model, the SN ejecta that hits the close companion is shocked and heats up.
The higher temperature gas emits UV and blue radiation in excess relative to an explosion where the ejecta does not undergo collision, or where the collision is at an unfavourable viewing angle.
We argue that while the observations can indeed be explained by shock interaction, it need not be with a companion in the SD scenario.
We present an alternative explanation, where the SN ejecta collides with material previously expelled in the WD-WD merger in the frame of the DD scenario.
\citet{Levanon2015} pointed out that when the ejecta collides with this disk-originated matter (DOM), both the ejecta and DOM heat up, hence adding to the radiation hours to days after explosion.
Such extra radiation, \citet{Levanon2015} argued, is in conflict with observations of SN~2011fe.
But it can account for excess radiation in the early light-curve when observed.

The paper is organized as follows.
In section \ref{sec:what-is-excess} we present a model-based definition for "excess light" and test whether the early light is indeed excessive for SN~2012cg.
We discuss the qualitative properties of the ejecta-DOM collision in section \ref{sec:dom-vs-sd-overview}, and compare to the collision of the ejecta with a companion.
In section \ref{sec:dom-vs-sd-models} we compute the expected luminosity from the ejecta-DOM collision and compare with observations of iPTF14atg and SN~2012cg.
We discuss our results and present our summary in section \ref{sec:summary}.

\section{Quantifying excess light in the early light curve}
\label{sec:what-is-excess}

Excess light is defined relative to a given model light curve.
For SN~2012cg, \citet{Marion2016} define excess light as deviation by more than $3\sigma$ from a power-law model $L \propto t^n$ fitted between $-14$ and $-8$ days from the time of B-band maximum light.
Power-law models provide adequate fits for the later part of many rising light curves \citep[e.g.][]{Firth2014} due to the fraction of \nickel\ at the diffusion depth being relatively constant.
This fit is not expected to hold for the early rising light curve, where the \nickel\ fraction at the diffusion depth is increasing \citep{Piro2014, Zheng2014}.
Since the shapes of early light curves vary, we instead define excess light to be light which cannot be accounted for by \nickel\ heating, regardless of light curve shape.
This means that the excess is relative to a more detailed progenitor and explosion model.
If the early observations of an event cannot be explained by the outermost \nickel\ in any of the available models, then this is a strong case for a different source of energy.
One potential source is shock cooling after the ejecta collides with matter in the SN vicinity, as elaborated on in the next sections.
We here demonstrate that early excess light can be seen in SN~2012cg when comparing to models for normal SNe~Ia.
This will in turn constrain the possible progenitor scenarios, though not exclusively to the SD scenario as we will show in section \ref{sec:dom-vs-sd-models}.

We use the formulation of \citet{Piro2013, Piro2014} to describe the early light curve.
The luminosity is a sum of two terms \Ldirect\ and \Ltail\ corresponding to heating above and below the diffusion depth, respectively:
\begin{equation}
\Ldirect = \int_0^t X_{56}(t') \frac{\partial \DMdiff}{\partial t'} \epsilon(t) dt',
\label{eq:L_direct}
\end{equation}
\begin{equation}
\Ltail = \int_t^{\tdiff(0)} X_{56}(t') \frac{\partial \DMdiff}{\partial t'} \epsilon(t)
	\frac{\erfc(t'/\sqrt{2}t)}{\erfc(1/\sqrt{2})} dt',
\label{eq:L_tail}
\end{equation}
where $X_{56}(t)$ is the \nickel\ fraction at the diffusion depth at time $t$ after the explosion, \DMdiff\ is the mass from the diffusion depth outwards, $\tdiff(0)$ is the diffusion time through the entire ejecta and $\epsilon(t)$ is the specific \nickel\ heating rate,
\begin{equation}
\epsilon(t) = \epsilon_{\rm Ni} e^{-t/t_{\rm Ni}}
	+ \epsilon_{\rm Co} \left( e^{-t/t_{\rm Co}} - e^{-t/t_{\rm Ni}}\right),
\label{eq:specific heating}
\end{equation}
with $\epsilon_{\rm Ni} = 3.9 \times 10^{10} \erggs$, $t_{\rm Ni} = 8.76 \days$, $\epsilon_{\rm Co} = 7.0 \times 10^9 \erggs$ and $t_{\rm Co} = 111.5 \days$.
We use a different method than \citet{Piro2013} to get the diffusion depth.
Following the explosion, the ejecta expands homologously, $r=v/t$.
We assume the ejecta has an exponential profile suitable for for SNe~Ia \citep{Dwarkadas1998},
\begin{equation}
\rho(t) = A e^{-v/\ve} t^{-3},
\label{eq:density}
\end{equation}
where
\begin{equation}
\ve = \left( \frac{E}{6\Mej} \right)^{1/2}, \quad
A = \frac{\Mej}{8 \pi \ve^3},
\label{eq:v_e and A}
\end{equation}
where $E$ is the explosion energy and \Mej\ is the ejecta mass.
The integral diffusion time from a radius $r$ outwards is \revision{approximated by}
\begin{equation}
t_{\rm diff}(r) = 2 \int_{r}^{\infty} \frac{3 \kappa \rho(t) r \cdot dr}{c}.
\label{eq:diffusion time}
\end{equation}
Integrating and equating $\tdiff(r)$ with the time from the explosion we can find the velocity at the diffusion depth $\vdiff(t)$ from the equation
\begin{equation}
t^2 = \frac{3 \kappa \Mej}{4 \pi c \ve} e^{-\vdiff/\ve} \left( \frac{\vdiff}{\ve} + 1 \right),
\label{eq:diffusion velocity}
\end{equation}
which gives the diffusion depth
\begin{equation}
\DMdiff = \Mej \left( \frac{4 \pi c \ve}{3 \kappa \Mej} t^2 + \frac{\vdiff}{2 \ve} e^{-\vdiff/\ve} \right),
\label{eq:diffusion depth}
\end{equation}
for $t < \tdiff(0)$.

To simulate the \nickel\ density profile we use a sigmoid function similar to \citet{Piro2014},
\begin{equation}
X_{56}(v) = \frac{X_{56}'}{1 + \exp\left[\beta(v - v_{1/2}) \right]}.
\label{eq:nickel profile}
\end{equation}
The parameters $(\beta,v_{1/2},X_{56}')$ are chosen to mimic the results of various explosion models with different \nickel\ mass and stratification.
The sigmoid profile, although simple, adequately models the sharp cut-off of the \nickel\ distribution, which is its most important property when modelling early light.

To create model light curves for normal SNe~Ia, we use values for \Mej, $E$, \MNi\ and parameters for $X_{56}(v)$ that bracket the results of simulations for various proposed progenitor models for normal SNe~Ia:
the delayed detonation of a Chandrasekhar-mass WD in the SD scenario \citep{Seitenzahl2013}, the DDet scenario \citep{Fink2010}, the violent merger in the DD scenario \citep{Pakmor2012} and the WD-WD collision scenario \citep{Kushnir2013}.
The parameter ranges used here are $\Mej=1.1-1.4\Msun,\, E=1-1.5\times10^{51}\erg$ and $\MNi=0.5-0.8\Msun$. $X_{56}(v)$ is chosen to have cut-offs with $X_{56}(6000-10000\kms)=0.5$ and $X_{56}(9000-11000\kms)=0.1$.
For the opacity we take $\kappa=0.05 \cmg$.
A constant opacity in the range $\kappa=0.03-0.1 \cmg$ is justified for the ejecta at 10 days past explosion \citep{Pinto2000}, and we use this simplifying assumption here.

Our model assumes the SN ejecta is spherically symmetrical, while most of the multidimensional numerical models quoted above show some level of asymmetry.
However, the main property governing the shape of the light curve is the stratified \nickel\ profile.
In an asymmetrical explosion there is some direction in which the \nickel\ is most outwardly extended.
We therefore choose the spherical profiles to give upper bounds on the location of the \nickel\ within the ejecta.
The goal is to see whether \nickel\ heating can be ruled out as the cause of the observed early light.

\begin{figure}
  \includegraphics[width = \columnwidth]{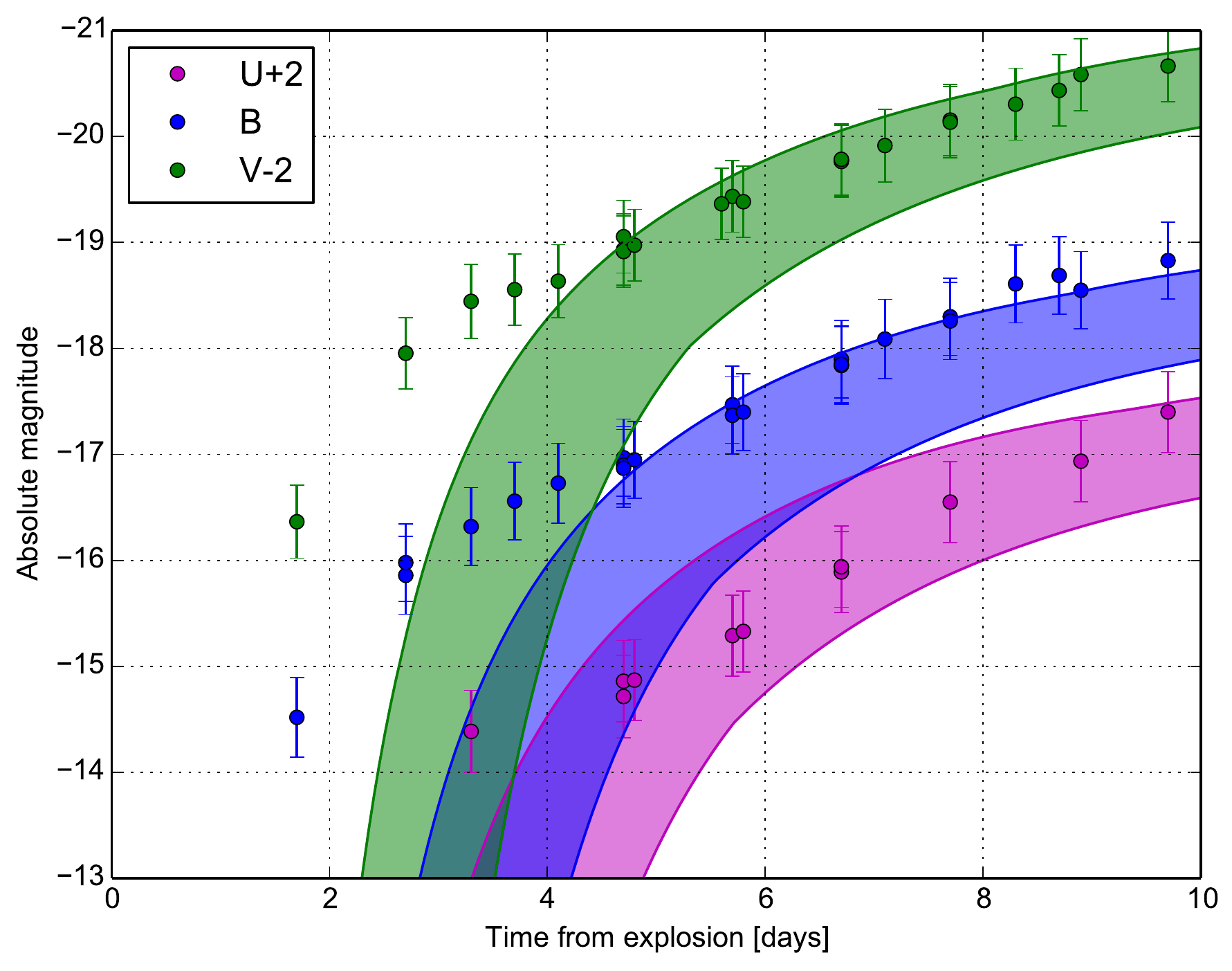}
  \caption{Model early light curves (shaded regions) and SN~2012cg observations (data points) from \citet{Marion2016}, using a rise time of 18.8 days to B-band maximum.
  Magnitudes are in the Vega system.
  The observations show excess light relative to what is expected from the common SN~Ia progenitor models due to their stratified \nickel\ structure.}
  \label{fig:SN2012cg vs models}
\end{figure}

Fig. \ref{fig:SN2012cg vs models} shows the resulting bounds on the early light curve from \nickel\ heating, with the observations of SN~2012cg taken from \citet{Marion2016}.
The shaded regions show the range of light curves obtained from the models.
The data points show the observations of SN~2012cg in the U, B and V bands in Vega magnitudes in purple, blue and green, respectively.
\revision{We assume a blackbody spectrum to generate the model light curves.
Consequently, the filtered light curves are upper limits if the color temperature is actually larger than the effective temperature \citep[e.g.][]{Rabinak2011}.}
The excess light in SN~2012cg can be seen for the early observations.
There is no parameter set for our model where \nickel\ heating can account for the early emission from SN~2012cg.
The time of explosion for SN~2012cg is in debate. \citet{Silverman2012}, \citet{Marion2016} and \citet{Shappee2016} find rise times of 17.3, 18.8 and 19.5 days to B-band maximum, respectively, without overlap in the errors.
If we take the most lenient value of 19.5 days, the excess is still seen for observations less than 4 days after the explosion.
We do not include the clear filter data from ROTSE as \citet{Marion2016} do by shifting it to the B-band, as this could present a spurious excess in flux instead of a different rise-time in the clear filter \citep{Shappee2016}.

\revision{The explosion time estimates were found using extrapolations of power-law fits to the light curve.
They do not include a possible ``dark phase" from explosion to first light due to the stratified \nickel\ structure \citep{Piro2013} as the models exhibit.
If the explosion occurs 20.5 days before B-band maximum, then the light curves are explained by the models and no excess light is seen.
The existence of early excess light in SN~2012cg therefore hangs on the assumption of a short dark phase of about one day or less.
We will assume that this is the case and explore how this possible excess light can be explained.
}

\begin{figure}
  \includegraphics[width = \columnwidth]{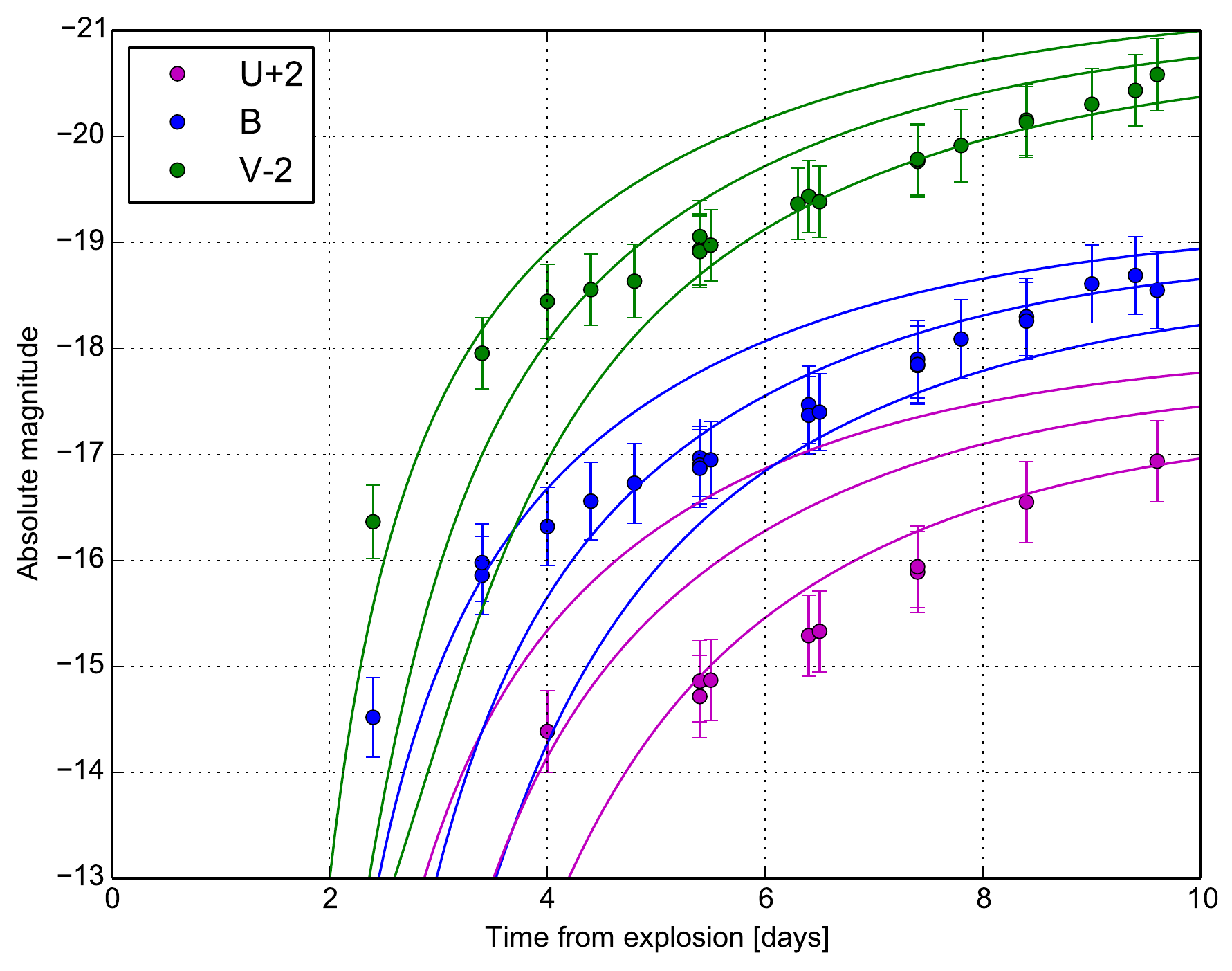}
  \caption{Same as Fig \ref{fig:SN2012cg vs models}, but with light curves for DDet models with an outer \nickel\ shell as solid lines. The models mimic the \nickel\ distribution of models 2-4 from \citet{Fink2010}, with an added shell of $X_{56}=0.05$ at higher velocities. In the plots model 2 is least luminous and model 4 is most luminous.}
  \label{fig:SN2012cg vs DDet models}
\end{figure}

The stratified \nickel\ structure which characterises progenitor models cannot account for the early light curve of SN~2012cg 
\revision{with the estimated explosion times}.
Therefore, models in which some of the \nickel\ lies close to the surface should be considered.
\citet{Piro2014} found that \nickel\ in SN~2012cg must lie at a depth $< 10^{-2} \Msun$ from the surface to explain the light curve, and suggested the DDet scenario as the first detonation produces \nickel\ at the surface of the WD.
Our results confirm those of \citet{Piro2014}, and so we examine the possibility of explaining the excess light with surface \nickel\ in the DDet scenario.
We use models with parameters as in models 2-4 of \citet{Fink2010} and add a shell of \nickel\ at velocities of $13700-14600\kms$ with an abundance of $X_{56}=0.05$ as an upper bound on the outer \nickel\ in the \citet{Fink2010} models.
We present our results for this model in Fig. \ref{fig:SN2012cg vs DDet models}.
While the outer \nickel\ does affect the light curve rise, it is not enough to explain the earliest observations.
We therefore confirm SN~2012cg to have excess light relative to normal SN~Ia models.
We consider a comparison to specific models to be a better criterion for defining excess (compared with assuming $L \propto t^n$) as it underlines that \nickel\ heating is insufficient to explain the observations and prompts us to add new ingredients to the models.

Applying this procedure to test the UV excess in the SN~Ia iPTF14atg is somewhat problematic, as it is a peculiar, sub-luminous SN, with an estimated \nickel\ mass of less than $0.2 \Msun$ \citep{Cao2016,Kromer2016}.
The UV data for iPTF14atg show a UV pulse declining at early times before rising again with the light curve.
While this does not rule out the possibility of a shallow \nickel\ shell, it is a feature which cannot be accounted for by a stratified \nickel\ structure.
We will for the next sections assume that it is not derived from shallow \nickel\ as was shown for SN~2012cg. 

\section{Differences between companion and DOM collisions}
\label{sec:dom-vs-sd-overview}

This section will highlight the differences between two SN ejecta collision models: ejecta-companion collision in the SD scenario \citep{Kasen2010} and ejecta-DOM collision in the DD scenario \citep{Levanon2015}.
An ejecta-companion collision can also take place in the DDet scenario, with the companion being a He-star.
In the DD scenario two WDs merge and subsequently explode as a SN~Ia.
The timing of the detonation is not agreed upon.
If the detonation occurs during the dynamical merger it is termed a violent merger \citep{Pakmor2012}.
At this stage the merger product is compact ($R<0.1\Rsun$), with the exception of the donor WD's tidal tail with a mass of $10^{-3}\Msun$ \citep{Raskin2013}.
If the detonation occurs hours after the merger during the viscous time-scale \citep[e.g.][]{vanKerkwijk2010}, then the donor WD forms an accretion disk \citep{Guerrero2004,Loren-Aguilar2009}.
Some of the accretion disk's material is expelled from the system due to angular momentum conservation.
We have termed this material DOM \citep{Levanon2015}.
\citet{Schwab2012,Schwab2016} argue that detonation conditions are not reached after merger and the merger burns stably until collapsing to form a neutron star.
Another alternative is that the merger remains over-stable due to rotation, and detonation condition are only reached after a delay of several million years \citep{Tornambe2013}.
If the merger does however explode on the viscous time scale, then the SN ejecta collides with the DOM and shock it.
The rest of the paper explores the consequences of the ejecta-DOM collision.

In the ejecta-DOM collision model the shocked material is comprised of donor WD material from the DOM and the outer parts of the SN ejecta.
The outer parts of the ejecta contain unburned carbon and oxygen and intermediate mass elements (IMEs) from the detonation's nucleosynthesis.
In contrast, in the ejecta-companion collision model the shocked material includes material sweeped by the ejecta from the non-degenerate companion.
The shocked matter will thus contain hydrogen or helium, depending on the companion type.
This difference may be seen in spectroscopy of SNe where an ejecta collision is suspected \citep[e.g.][]{Maguire2016}.

\revision{We assume that a mass of about $0.01-0.1 \Msun$ is expelled in the DOM.
In contrast, \citet{Schwab2012} found in their simulations of merging WDs that only a mass of $ \la 10^{-5} \Msun$ is unbound in the merger.
We attribute this large difference to the purely hydrodynamic nature of their simulations.
\citet{Ji2013} simulated the viscous evolution of a WD merger with magnetic fields.
They find outflows of disk material to comprise nearly $0.06 \Msun$, and while most of this mass remains gravitationally bound, it reaches large radii (outside their simulation domain) before falling back in.
It is thought that strong magnetic fields are responsible for the launching of about 1--10 per cent of the gas that is accreted by accretion disks in young stellar objects and active galactic nuclei.
In young stellar objects the jets might carry about 10--40 per cent of the mass that flows through the accretion disk \citep{Pudritz2012,Federrath2014}.
Based on the behavior of jets and magnetic fields in these well studied objects, we assume that several per cent of the destructed WD are launched in the DOM.}

In the ejecta-DOM collision model, the DOM is moving outwards at a speed of $\approx 5000 \kms$ when the ejecta hits it at a speed of $\approx 10000 \kms$.
In the ejecta-companion collision however the companion is stationary so that the relative impact speed is double.
The pressure after the shock is then four times larger than for an ejecta-DOM collision.
However this does not take into account the structure of the shocked region which will also influence its pressure profile.
The pressure gradient then determines the shape and power of the shock cooling light curve.
In this paper a simple model for the shocked DOM structure is assumed to test the viability of the ejecta-DOM collision model for explaining early excess light.

The time and radius of the ejecta-DOM collision are set by the delay time between the onset of accretion and the explosion.
The delay also determines the mass of the DOM as it is continually expelled during the accretion.
Since the ejecta speed is about double the DOM speed, the collision time will be roughly as the delay time before explosion $\approx 10^4 \s$.
The collision radius is then $\approx 100\Rsun$.
If the delay is longer, the collision might take place at larger distances up to several AU.
The excess emission in that case appears later, hours to days after explosion.
For the ejecta-companion collision the radius is set by the binary separation during mass transfer.
The collision radius can be between about $4\Rsun$ for a $1\Msun$ main sequence companion and about $300\Rsun$ for a red giant companion.
A small progenitor radius, if observed, can thus constrain the possibility of an ejecta collision \citep[e.g.][]{Bloom2012}.

Both collision models have viewing angle dependence.
The shock interaction is seen only when the collision is at a favourable viewing angle, i.e., towards the observer.
The DOM is expelled either in the polar directions as jets or in the equatorial plane as a result of the tidal destruction.
We estimate the DOM should cover 10 to 30 per cent of the sphere.
This is similar to the spherical coverage of a companion, and so only for about 10 per cent of events should the interaction be seen for both collision models \citep{Kasen2010}.

\section{The ejecta-DOM collision}
\label{sec:dom-vs-sd-models}

The derivation of early SN~Ia light curves from an ejecta-DOM collision follows the calculations in section 4 of \citet{Kasen2010}.
In those calculations the luminosity of the SN ejecta is computed under the assumption that part of the ejecta was previously shocked when it collided with the companion.
The ejecta profile used by \citet{Kasen2010} is a broken power-law \citep{Chevalier1989}, and the luminosity is calculated in the diffusion approximation \citep{Chevalier1992}.
As an alternative to the process of shocking the ejecta on a stellar companion, we propose that in the DD scenario the ejecta might collide with the DOM, as described in section \ref{sec:dom-vs-sd-overview}.
While the power-law profile gives an analytic self-similar solution, it is more suitable for SN~II ejecta.
We instead use the exponential density profile from equation \ref{eq:density}.
The ejecta moves into the DOM, of a mass of $\Mdom \approx 0.01-0.1 \Msun$, which was expelled $\Dtexp \approx 10^4 \s$ prior to the explosion at a speed of $\vdom \approx 5000 \kms$.
The DOM is expelled primarily in the polar directions due to jets (or a wind from the disk surface), and we will assume it covers a fraction $\fdom \approx 0.1$ of the sphere.
When the ejecta and the DOM collide, the mass of the ejecta that suffers a strong shock before the DOM is accelerated is several times the mass of the DOM.
The shock wave moves through the DOM very swiftly, resulting in a transient signal \citep{Levanon2015}, and afterwards the entire DOM is shocked, along with some ejecta material.
This strong interaction ends approximately at a time \ts\ and velocity coordinate $\vs = \rs/\ts$ where the shocked DOM and outer mass of the ejecta reach equal speed.
From conservation of momentum,
\begin{multline}
\frac{\Mdom}{\fdom} \vdom +
4\pi \int_{\vs}^\infty \! A e^{-v/\ve} v^3 \mathrm{d}v \\
= \left( \frac{\Mdom}{\fdom} +
4\pi \int_{\vs}^\infty \! A e^{-v/\ve} v^2 \mathrm{d}v \right) \vs.
\label{eq:shock radius}
\end{multline}
For $E = 10^{51} \erg$, $\Mej = 1.4 \Msun$, $\Mdom = 0.05 \Msun$ and $\Dtexp = 2 \times 10^4 \s$ this gives $\vs \approx 10600 \kms$, and the total strongly shocked mass is $\Ms = 6.4 \Mdom$.
Since the DOM was expelled earlier, its radius at the end of the interaction is
\begin{equation}
\rs \approx \vdom \left( \ts + \Dtexp \right).
\label{eq:r_dom}
\end{equation}
With $\rs = \vs \ts$ this gives $\rs = 270 \Rsun$ and $\ts = 0.2 \days$.

The outer ejecta and the DOM are shocked to very high temperatures, $T > 5 \times 10^5 \K$, such that radiation pressure dominates and we can take $\gamma = 4/3$.
We will assume that the profile of the shocked material remains exponential and is compressed by a factor $\fcomp$ to $\rhos = \As e^{-v/\ve\fcomp} \ts^{-3}$ (compare with equation \ref{eq:density}).
The scaling of the shocked profile is found via conservation of mass to be
\begin{equation}
\As = A \frac{\Ms}{\Mej} 
\left[ \fcomp^3 e^{-\vsnorm} \left( 1 + \vsnorm + \frac{1}{2} \left(\vsnorm\right)^2 \right) \right]^{-1}.
\label{eq:A_s}
\end{equation}
As in \citet{Kasen2010}, the initial pressure of the shocked material is the ram pressure $\ps = \frac{1}{2} \rhos \left( \vs - \vdom \right)^2$.
The pressure decreases adiabatically with time as $\pf = \ps \left(  \rhof/\rhos \right)^\gamma$, where $\rhof = \rhos (t/\ts)^{-3}$.
We compute the luminosity under the diffusion approximation.
The diffusion wave radius in the shocked material is found via equation \ref{eq:diffusion time} with the time from collision $t-\ts$ on the left hand side.
Since the shocked region in relatively thin we do not consider the diffusive tail as in section \ref{sec:what-is-excess}.
As in \citet{Kasen2010}, the comoving luminosity in the diffusion approximation with radiation pressure is computed from the pressure gradient at the diffusion radius \citep{Chevalier1992},
\begin{multline}
L_{\rm shock} =
\left. -4 \pi \fdom r^2 \frac{c}{\kappa \rhof}
\frac{\partial \pf}{\partial r} \right|_{r = \rdiff} \\
= \frac{2 \pi \fdom c}{\kappa \ve \fcomp} \left(\vs-\vdom\right)^2 \ts \vdiff^2.
\label{eq:L_shock}
\end{multline}
To compute \Teff\ we find the photospheric radius \rph\ from $\int_{\rph}^{\infty} \kappa \rhof dr = 2/3$ to be
\begin{equation}
\rph = \ve \fcomp \ln\left( \frac{3 \As \kappa}{2 t^2} \ve \fcomp \right) t,
\label{eq:r_ph}
\end{equation}
and the effective temperature is $\Teff = [ L / (4 \pi \fdom \rph^2 \sigma) ]^{\frac{1}{4}}$.
The decrease in luminosity due to taking a partial coverage of $4 \pi \fdom$ is not seen in $\Teff$ but is taken into account when computing the magnitudes.
After all the shocked material becomes diffusive, the luminosity decreases faster, as the underlying layers are only weakly shocked.
We do not compute the luminosity for this stage.

\begin{figure}
\includegraphics[width = \columnwidth]{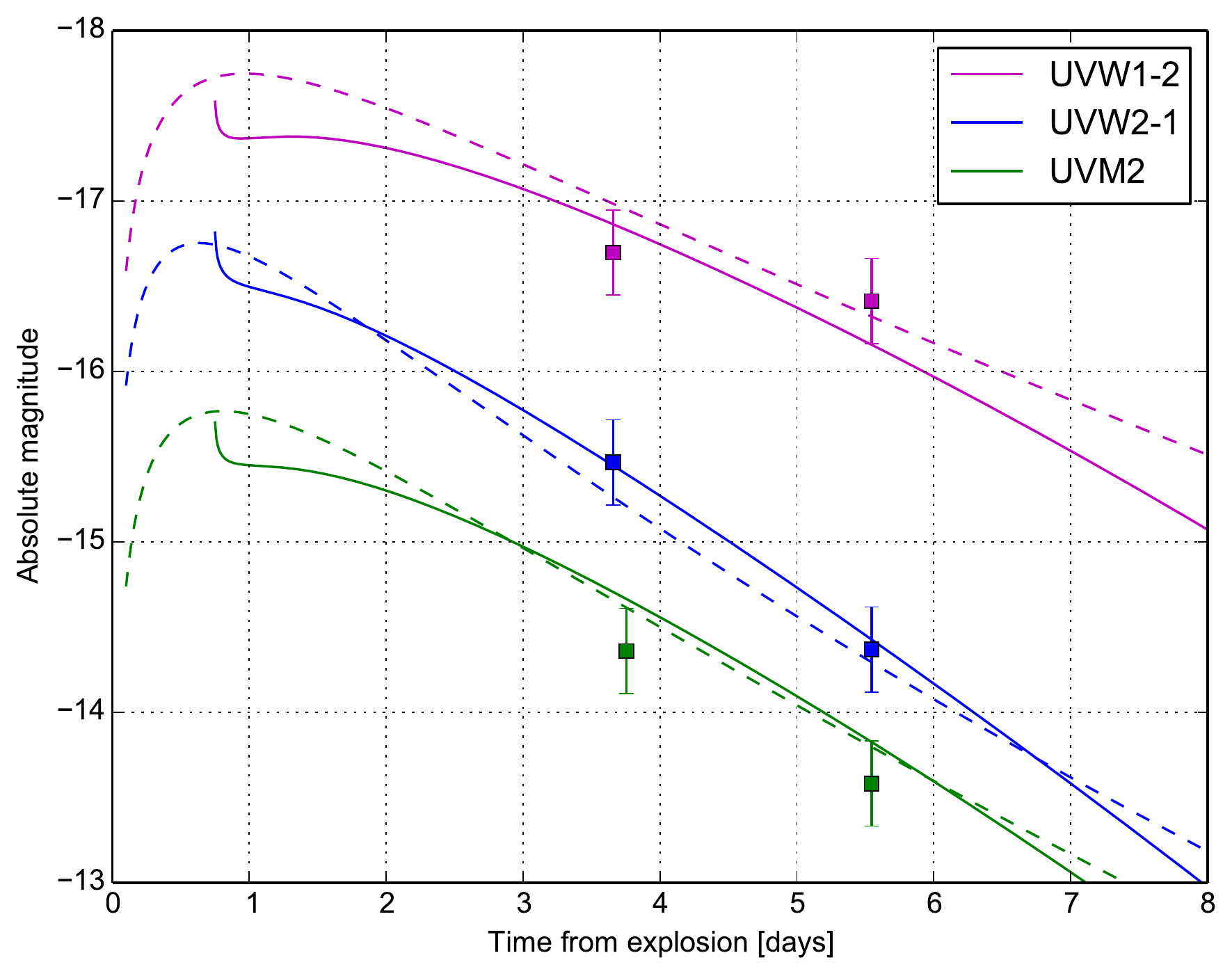}
\caption{Comparison of the effects of ejecta colliding with material in its vicinity in the DOM and SD models, with UV data of iPTF14atg from \citet{Cao2015}.
DOM model in solid curves with $\Mdom=0.03 \Msun$, $\Dtexp = 1.7 \times 10^4 \s$, $\fdom=0.1$, $\fcomp=1.5$,
SD model in dashed curves with $a=70 \Rsun$.
Both with $E=0.3 \times 10^{51} \erg$, $\Mej=1.4 \Msun$, $\kappa = \kappa_e$.}
\label{fig:DOM vs SD for iPTF14atg}
\end{figure}

Fig. \ref{fig:DOM vs SD for iPTF14atg} shows a comparison between our DOM model and the SD model used in \citet{Cao2016} for iPTF14atg.
The SD collision model in dashed curves has $E=0.3 \times 10^{51} \erg$, $\Mej=1.4 \Msun$ and a companion separation of $a=70 \Rsun$.
The DOM collision model in solid curves has $E=0.3 \times 10^{51} \erg$, $\Mej=1.4 \Msun$, $\Mdom=0.03 \Msun$, $\Dtexp = 1.7 \times 10^4 \s$, $\fcomp=1.5$.
The relatively low explosion energy in the DOM collision model is chosen to accord with the values chosen for the companion collision model by \citet{Cao2016}.
iPTF14atg is a peculiar, subluminous SN~Ia so that an explosion energy below $10^{51} \erg$ is justified.
The \textit{Swift} UV observations of the early pulse in iPTF14atg are shown in AB magnitudes as data points.

\begin{figure}
\includegraphics[width = \columnwidth]{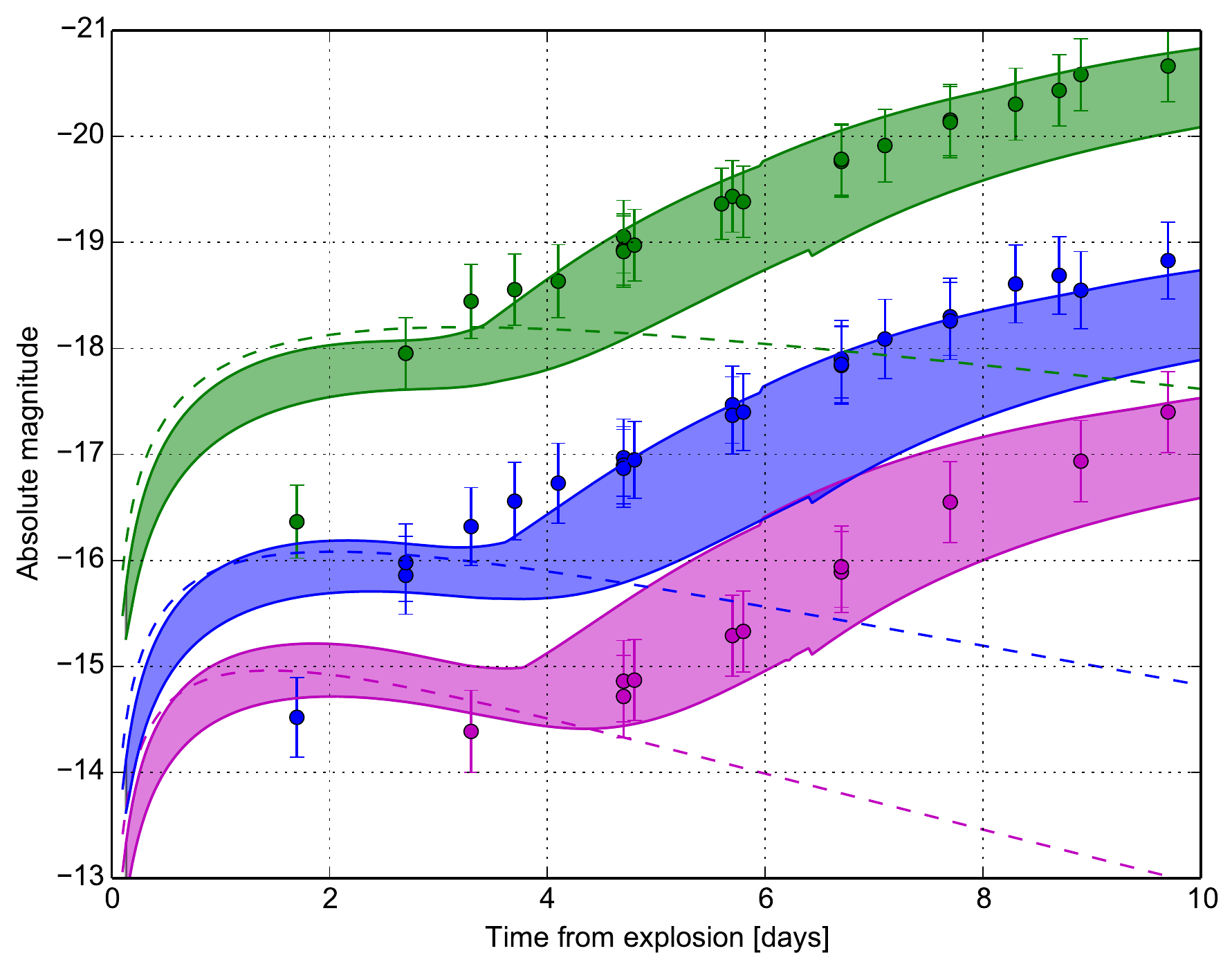}
\caption{Comparison of the effects of ejecta colliding with material in its vicinity in the DOM and SD models, with optical data of SN~2012cg from \citet{Marion2016}.
DOM models in filled regions with $E=1-1.5 \times 10^{51} \erg$, $\Mej=1-1.4 \Msun$, $\Mdom=0.1 \Msun$, $\kappa=0.05 \cmg$, $\Dtexp = 0.5 \times 10^4 \s$, $\fdom=0.1$, $\fcomp=1.5$,
SD model in dashed curves with $E=10^{51} \erg$, $\Mej=1.4 \Msun$, $a=2 \times 10^{12} \cm$, $\kappa = \kappa_e$.
Light from \nickel\ heating is added to the DOM models.
Explosion time is 18.8 days before B-band maximum \citep{Marion2016}.
Colors are the same as in Figs. \ref{fig:SN2012cg vs models} and \ref{fig:SN2012cg vs DDet models}}
\label{fig:DOM vs SD for SN2012cg 18.8d}
\end{figure}

For SN~2012cg \nickel\ heating should be noticeable in all but the earliest observations regardless of the specific time of explosion, as seen in Figs. \ref{fig:SN2012cg vs models} and \ref{fig:SN2012cg vs DDet models}.
Modelling its early light curve therefore requires adding the light from \nickel\ heating to the light from shock heating.
Fig. \ref{fig:DOM vs SD for SN2012cg 18.8d} shows the combined DOM and \nickel\ heating models for SN~2012cg. The SD collision model \citet{Marion2016} in dashed curves has $E=10^{51} \erg$, $\Mej=1.4 \Msun$ and a companion separation of $a=2 \times 10^{12} \cm$ for a $6\Msun$ main sequence companion.
Our DOM collision model has $\Mdom=0.1 \Msun$, $\kappa=0.05 \cmg$, $\Dtexp = 0.5 \times 10^4 \s$, $\fdom=0.1$, $\fcomp=1.5$.
A drop in luminosity is seen for models with \nickel\ only below $9000\kms$, where the luminosity from \nickel\ heating only rises later.
This is a synthetic artefact due to the luminosity from the ejecta-DOM collision being cut off in the model once the diffusion depth recedes behind the strongly shocked region.
In reality the drop in luminosity from the collision is gradual, and this effect can be disregarded in the graphs.

\begin{figure}
\includegraphics[width = \columnwidth]{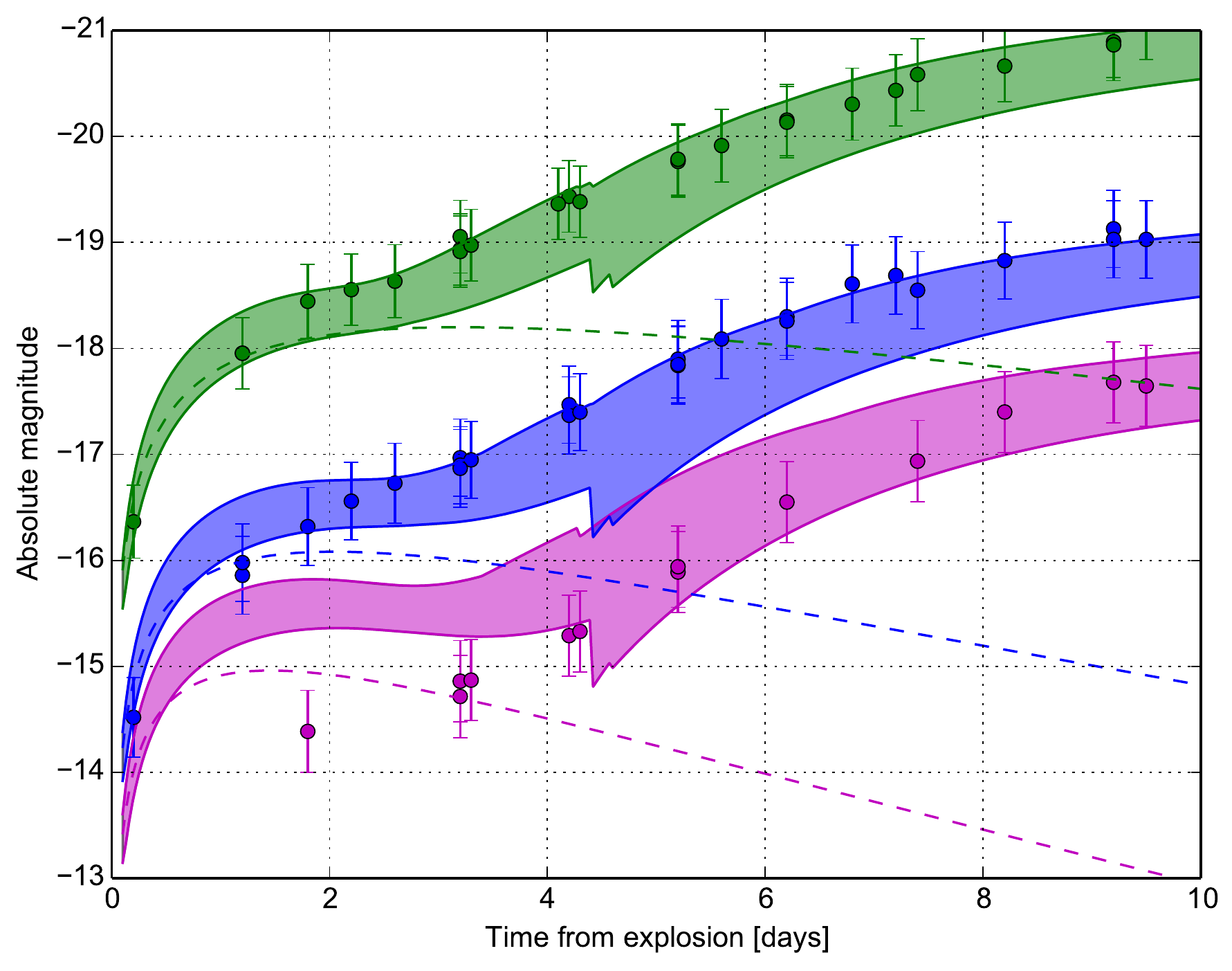}
\caption{Same as Fig. \ref{fig:DOM vs SD for SN2012cg 18.8d} with explosion time 17.3 days before B-band maximum \citep{Silverman2012}.
DOM models in filled regions with $E=1-1.5 \times 10^{51} \erg$, $\Mej=1-1.4 \Msun$, $\Mdom=0.1 \Msun$, $\kappa=0.03 \cmg$, $\Dtexp = 0.5 \times 10^4 \s$, $\fdom=0.15$, $\fcomp=1.5$,
SD model in dashed curves with $E=10^{51} \erg$, $\Mej=1.4 \Msun$, $a=2 \times 10^{12} \cm$, $\kappa = \kappa_e$.
Light from \nickel\ heating is added to the DOM models.}
\label{fig:DOM vs SD for SN2012cg 17.3d}
\end{figure}

Taking a rise time of 18.8 days as in \citet{Marion2016}, we find that the earliest B and V-band observations (from KAIT) are too dim to be fitted by the DOM model.
This is seen also for the SD collision model in \citet{Marion2016}.
Taking instead a shorter rise time of 17.3 days from \citet{Silverman2012}, the early observations are explained by both collision models, but the models are too bright in the U-band.
Additionally, both DOM and SD collision models are too bright in the \textit{Swift} UV bands, regardless of the choice of explosion time.
This suggests that a collision model may not be appropriate for SN~2012cg, or else a more detailed collision structure must be developed that can explain the relatively dim UV observations.

\section{Discussion and summary}
\label{sec:summary}

In this paper we explored the possibility of interpreting early excess light in SNe~Ia as an ejecta-DOM interaction in the DD scenario.
In section \ref{sec:what-is-excess} we presented a definition of "excess light" which is relative to the light curves of potential explosion models.
We see this as a more physically grounded definition than comparing to a power-law model as done by \citet{Marion2016}.
This method confirms previous findings that SN~2012cg had an early excess blue light.
We presented our results in Figs. \ref{fig:SN2012cg vs models} and \ref{fig:SN2012cg vs DDet models}.
The early excess light observed in some SNe~Ia is still a puzzle.
In the present paper we proposed that this excess emission, mainly in the UV, can be accounted for within the frame of the DD scenario.
Our study was motivated in part by the problems that other explanations have encountered.
We discuss some of these problems in this section.

The source of the early excess light in the SNe~Ia iPTF14atg and SN~2012cg is explored in several papers.
\citet{Kromer2016} compare the light curves and spectra of iPTF14atg to multidimensional explosion models.
They find that explosion models in the SD scenario do not fit the observations of iPTF14atg.
Their delayed detonation models produce too much \nickel\ which leads to larger peak luminosities, and their deflagration models have a faster time evolution of light curves and spectra compared to iPTF14atg.
\citet{Kromer2016} find better agreement for a violent merger in the DD scenario, between WDs of masses $0.9\Msun$ and $0.76\Msun$ for observations 10 days or later after the explosion.
The UV pulse is not explained by the violent merger.
The explosion in this scenario takes place on the dynamical time-scale of the WD-WD merger, while the accretion disk forms and before the DOM is expelled.
Our proposed ejecta-DOM collision model can work only if the explosion is somehow delayed to the viscous time-scale.
If there is no delay between merger and explosion, the only material expelled will be the donor WD's tidal tail \citep{Raskin2013}.
When the tidal tail is hit by the explosion ejecta, the light from shock cooling will be weaker compared to that of an ejecta-DOM collision, as the tidal tail's mass is only $\sim10^{-3}\Msun$.

\citet{Liu2015} use binary population synthesis and the mechanism studied by \citet{Kasen2010} of ejecta-companion collision within the frame of the SD scenario.
They thus estimate the early UV light expected for different SD progenitor systems.
They find that He-star and main sequence donors do not provide enough light when the ejecta collides with them to account for the UV pulse of iPTF14atg and that a red giant companion is required to account for the UV pulse.
A red giant companion would have larger separation of about $a \approx 10^{13}\cm$, however, which means that to get the same luminosity as the collision model of \citet{Cao2016} a lower explosion energy of $E < 10^{50} \erg$ is required.

\citet{Cao2016} compare iPTF14atg with another SN~2002es-like event, iPTF14dpk.
They find that iPTF14dpk does not contain any evidence of early excess light, and conclude that if both SNe are indeed from the same progenitor system then this is a viewing angle effect.
Both a companion and DOM should have a viewing angle dependence (see section \ref{sec:dom-vs-sd-overview}).
We agree that shock interaction of the ejecta with material near the SN~Ia is a likely source for the UV pulse of iPTF14atg considering its timing and wavelength.
However we find our proposed ejecta-DOM interaction to better account for this UV excess, considering the difficulties with the SD scenario for this event listed above.
We presented our model's fit to the UV excess in Fig. \ref{fig:DOM vs SD for iPTF14atg}.

In the SD scenario all SNe~Ia have UV excess, but its visibility depends on the viewing angle.  
In the DD scenario there is a possibility that if there is no delay between merger and explosion, or if the delay is very long, the explosion will not have a UV excess.
In our model in addition to the viewing angle, the delay time between merger and explosion is also a parameter to determine whether there is a UV excess or not.

\citet{Shappee2016} argue against a SD scenario for SN~2012cg.
They present several different observations that are at odds with a non-degenerate companion.
In particular, pre-discovery observations limit the progenitor radius to $R<0.23 \Rsun$ for most viewing angles, including those for which ejecta-companion interaction should be seen.
While we have shown that SN~2012cg does have excess blue light relative to explosions with stratified \nickel\ structure, the interaction models do not provide a full solution.
As seen in Figs. \ref{fig:DOM vs SD for SN2012cg 18.8d} and \ref{fig:DOM vs SD for SN2012cg 17.3d}, both interaction models over-estimate the early observations by about one magnitude in some of the filters (and this is true for the UV filters as well).
We conclude that a more detailed handling of the shocked material's structure and opacity is required for either collision model, or else an explosion with more surface \nickel\ must be invoked to explain these observations \citep[e.g.][]{Piro2016}.
In light of the progenitor radius constraint of \citet{Shappee2016}, surface \nickel\ seems a more likely explanation.

To summarize, we have shown that the ejecta-DOM interaction models can fit equally well as the ejecta-companion interaction models for iPTF14atg (Fig. \ref{fig:DOM vs SD for iPTF14atg}) and SN~2012cg (Figs. \ref{fig:DOM vs SD for SN2012cg 18.8d} and \ref{fig:DOM vs SD for SN2012cg 17.3d}).
The ejecta-DOM collision provides an explanation for the early excess light in the frame of the DD scenario. 
We view the DD scenario as more likely for these events given the constraints on a non-degenerate companion from other observations \citep{Kromer2016,Shappee2016}.

We emphasize the importance of considering all scenarios when observations are compared to theory.
Finding that a model of a certain scenario fits observations does not rule out that models of a different scenario, or just of a different variant of the same scenario, may fit as well.

In all five SN~Ia scenarios compared by \citet{Tsebrenko2015} and \citet{Soker2015}, the WD explodes after interacting with a stellar companion.
It is not surprising then that some properties of SNe~Ia are shared by two or more of these scenarios.
We have shown here that early excess light can be explained in the DD scenario as an ejecta-DOM collision, just as well as an ejecta-companion collision in the SD scenario.
To constrain the possible scenarios further, more detailed modelling is required.

\section*{Acknowledgements}

\revision{We thank an anonymous referee for helpful comments.}
This research was supported in part by a grant from the Israel Science Foundation.
This research made use of Astropy, a community-developed core Python package for Astronomy \citep{Robitaille2013}.\footnote{\url{http://www.astropy.org}}



\bibliographystyle{mnras}
\bibliography{Mendeley}


\bsp	
\label{lastpage}
\end{document}